\documentclass[preprint,12pt]{elsarticle}
\biboptions{sort&compress}

\usepackage{amssymb}
\usepackage[T1]{fontenc}
\usepackage{amsmath}
\usepackage{placeins}
\usepackage{tabularray}
\usepackage{xcolor}
\usepackage{multirow}
\usepackage{booktabs}
\usepackage{caption}
\usepackage{subcaption}
\usepackage{listings}
\definecolor{codegreen}{rgb}{0,0.6,0}
\definecolor{codegray}{rgb}{0.5,0.5,0.5}
\definecolor{codepurple}{rgb}{0.58,0,0.82}
\definecolor{backcolour}{rgb}{0.95,0.95,0.92}

\lstdefinestyle{mystyle}{
    backgroundcolor=\color{backcolour},   
    commentstyle=\color{codegreen},
    keywordstyle=\color{magenta},
    numberstyle=\tiny\color{codegray},
    stringstyle=\color{codepurple},
    basicstyle=\ttfamily,
    breakatwhitespace=false,         
    breaklines=true, 
    captionpos=t,    
    keepspaces=true, 
    numbers=left,    
    numbersep=5pt,  
    showspaces=false, 
    showstringspaces=false,
    showtabs=false,  
    tabsize=2
}

\newcommand{\R}{\textsuperscript{\textregistered} }

\newcommand{\edit}[1]{\textcolor{black}{#1}}
\newcommand{\PoF}[1]{\textcolor{black}{#1}}

\journal{Physics of Fluids}

\begin{document}

\begin{frontmatter}



\title{\edit{Assessing nozzle flow dynamics in Fused Filament Fabrication through the parametric map $\alpha-\lambda$}}


\author[inst2,inst3,inst4]{Tom\'as Schuller}
\author[inst6]{Paola Fanzio }
\author[inst4,inst5]{Francisco J. Galindo-Rosales \corref{cor1} }

\cortext[cor1]{Corresponding author:~\edit{galindo@fe.up.pt}}

\affiliation[inst2]{organization={Institute of Science and Innovation in Mechanical and Industrial Engineering (INEGI)},
            addressline={Rua Dr. Roberto Frias, 400}, 
            city={Porto},
            postcode={4200-465},
            country={Portugal}}

\affiliation[inst3]{organization={Transport Phenomena Research Center (CEFT), Mechanical Engineering Department},
            addressline={Faculty of Engineering of the University of Porto, Rua Dr. Roberto Frias s/n}, 
            city={Porto},
            postcode={4200-465}, 
            country={Portugal}}

\affiliation[inst4]{organization={ALiCE—Associate Laboratory in Chemical Engineering},
            addressline={Faculty of Engineering of the University of Porto, Rua Dr. Roberto Frias s/n}, 
            city={Porto},
            postcode={4200-465}, 
            country={Portugal}}

\affiliation[inst6]{organization={Department of Precision and Microsystems Engineering (PME), Faculty of Mechanical, Maritime and Materials Engineering (3mE)},
            addressline={TU Delft (Delft University of Technology), Mekelweg 2}, 
            city={Delft},
            postcode={2628 CD}, 
            country={The Netherlands}}

\affiliation[inst5]{organization={Transport Phenomena Research Center (CEFT), Chemical Engineering Department},
            addressline={Faculty of Engineering of the University of Porto, Rua Dr. Roberto Frias s/n}, 
            city={Porto},
            postcode={4200-465}, 
            state={},
            country={Portugal}}            
            
\begin{abstract}

\edit{Polymer rheology profoundly influences the intricate dynamics of material extrusion in Fused Filament Fabrication (FFF). This numerical study, which uses the Giesekus model fed with a full rheometric experimental data set, meticulously examines the molten flow patterns inside the printing nozzle during FFF. Our findings reveal new insights into the interplay between elastic stresses and complex flow patterns, highlighting their substantial role in forming upstream vortices. The parametric map $\alpha$-$\lambda$ from the Giesekus model allowed us to sort the materials and connect the polymer rheology with the FFF nozzle flow dynamics. The identification of elastic instabilities, the characterization of flow types, and the correlation between fluid rheology and pressure drop variations mark significant advancements in understanding FFF processes. These insights pave the way for tailored nozzle designs, promising enhanced efficiency and reliability in FFF-based additive manufacturing.}

\end{abstract}





\end{frontmatter}


\FloatBarrier
\section{Introduction}

Material extrusion, an additive manufacturing (AM) method, entails precisely dispensing material through a nozzle, following established standards \cite{ISO52900}. Fused filament fabrication (FFF), introduced by S. Scott Crump in 1989 \cite{PatentCrump}, currently stands as the predominant technology for material extrusion \cite{NAJMON20197}. FFF allows for the rapid, cost-effective creation of intricate three-dimensional objects suitable for practical applications \cite{Klippstein}. Moreover, post-processing techniques like sanding, chemical treatment, polishing, \edit{and} metal plating, among others, can be employed to enhance surface quality \cite{LALEHPOUR201642,polym11030566,Kristiawan2021}.

The FFF procedure consists of introducing thermoplastic materials, usually in filament shape, from spools into a temperature-regulated liquefier, where they \edit{transform into} liquid state. Subsequently, the molten material is extruded through a nozzle and guided by the print head onto a build plate in specific patterns to create a 3D object. Common quality issues in FFF systems encompass over-extrusion and under-extrusion, leading to the formation of blobs or holes. Another extrusion challenge, known as annular backflow, happens when the molten polymer reverses its flow along the ring-shaped region between the filament and the liquefier wall, escaping the heated area and cooling below the glass temperature, where solid/fluid transition occurs ($T_g$) \cite{gilmer}. These issues undoubtedly arise from the interplay of extrusion process parameters with the rheological properties of the molten filament \cite{Arit2021}\edit{; however, state-of-the-art lacks fundamental research to understand the underlying mechanisms fully that allow for technical solutions to these issues.} 

\edit{The printing nozzle is at the heart of the FFF process, being a critical component responsible for precise material deposition~\cite{OSSWALD201851,GAO2021101658,Yadav2022}. The nozzle is a metal piece connecting the liquefier to the die through a contraction.} It is well documented \edit{in} the literature~\edit{\cite{McKinley1991,oliveira1,Baaijens1993,rothstein1,rothstein,Pimenta2020,POOLE2023105106}} that the flow of viscoelastic fluids through an axisymmetric contraction geometry can lead to elastic instabilities responsible for generating not only upstream vortices but also excess pressure drop. Furthermore, smaller characteristic length scales in the geometry intensify elastic effects. Previous efforts sought to replicate the polymer melt extrusion process within an FFF printer nozzle using microfluidic chips of a planar geometry with shear-thinning polymer solutions. This led to the construction of a Deborah-Reynolds flow map, revealing the production of growing upstream vortices under typical printing flow conditions in an FFF nozzle die \cite{rooooooooy}. Subsequently, the formation of vortices upstream of the tapered region in the nozzle, driven by elastic instabilities, was observed. This study was complemented in \cite{schuller_add_ma}, where numerical analysis confirmed the presence of these vortices when molten Polycarbonate flowed through an axisymmetric nozzle geometry. It also allowed for the disentanglement of the relative significance of extension-induced and shear-induced elastic stresses in this extrusion process. It was concluded that within the upstream vortex, extensional elastic stresses predominated, but their importance diminished with higher extrusion velocities; moreover, they played a determinant role in the size of the upstream vortex structures. However, shear-induced normal stress differences may account for the excess pressure drop and a change in equilibrium height ($H^*$) in the backflow region, where the polymer melt flows upward between the solid filament and the liquefier wall.
Recently, an experimental study with \edit{Polylactic Acid (PLA)} \cite{sietsepaper} allowed the experimental determination of the nozzle pressure drop, which, in combination with traditional top-of-nozzle pressure measurements, permitted determining the pressure drop occurring in the liquefier and the calculation of equilibrium height in the backflow region based on experimental data exclusively. Results showed a non-linear increase in the pressure measured by the feeders, which was associated with shear-induced elastic instabilities occurring in the backflow region at the same extrusion velocity at which the maximum equilibrium height ($H^*$) occurred. 

\edit{Most of the research papers available in the literature dealing with FFF are focused either on a single polymeric material, such as our previous works \cite{schuller_add_ma, sietsepaper}, or on exclusively the viscous effect on flow dynamics~\cite{bellini,SERDECZNY2020101454,Mishra2022}, disregarding the viscoelastic intrinsic nature of the polymer melts affecting the stability of the flow through the nozzle at real printing speed~\cite{rooooooooy}. In this work, we aim to go beyond the state-of-the-art by analyzing numerically how the non-linear rheological properties of different materials commonly used in FFF, such as PET-G, PET-CF, PLA, PC, ABS and PA6/66, can be reduced to limited numbers of parameters, i.e. Giesekus model parameters, and how this can help in sorting the materials and understanding the FFF nozzle flow dynamics; special attention will be given to the shear and extensional stress developed in the fluid, and how these are connected elastic instabilities, extra pressure drop and, ultimately, backflow issues}. 

\FloatBarrier
\section{ Materials and Methods}

\subsection{\edit{Selection, characterization and modeling of the polymers}}

Polymer choice is pivotal in FFF 3D printing, directly influencing the produced part characteristics and performance. Common thermoplastics like ABS, PET-G, and PLA offer ease of use, while engineering-grade polymers such as Nylon and Polycarbonate enhance strength and temperature resistance. Polymers and composites with additives like carbon fiber or metal particles offer unique properties like conductivity or flame resistance. Selecting the most suitable polymer depends on specific application needs, considering factors like mechanical load, environment, and functionality. Filament diameter, quality, and printer compatibility are also important considerations.\\
The polymers that have been selected for this study are:

\begin{enumerate}
    \item PET-G: \edit{It is a} glycol-modified thermoplastic polyester \edit{that} offers durability, chemical resistance, and formability. It surpasses PLA in strength, impact resistance, and temperature tolerance, making it a versatile choice \cite{LatkoDuraek2019}. UltiMaker's transparent PET-G was the material of choice in this study.

    \item PLA: Derived from renewable sources, PLA is an environmentally friendly thermoplastic. It is cost-efficient and widely used in 3D printing, food, and medical industries \cite{Kristiawan2021,swetham2017critical}. In this analysis, as in previous work \cite{sietsepaper}, UltiMaker transparent PLA was used.

    \item PC: Polycarbonate is a strong thermoplastic known for impact resistance, heat tolerance, and fire retardancy, making it suitable for challenging environments \cite{krache2011some}. For this study, as in \cite{schuller_add_ma}, UltiMaker transparent PC was used.

    \item PA6/66: A copolymer of PA6 and PA66, PA6/66 offers high rigidity, heat resistance, and chemical resistance. It is ideal for industrial applications, including automotive and aerospace \cite{bierogel2014materials,bierogel2014quasi}. DSM Novamid\R ID 1030 (PA6/66) black was utilized in this investigation.

    \item ABS: ABS is an impact-resistant engineering thermoplastic commonly used in structural applications and various products \cite{Kristiawan2021,krache2011some}. In this study, the choice of material fell on UltiMaker white ABS.

    \item PET-CF: PET-CF, a carbon fiber-reinforced PET variant, combines printability with enhanced stiffness, chemical, and temperature resistance, making it suitable for industrial 3D printing \cite{sharma2021effect}. For this research LUVOCOM\R 3F PET-CF 9780 Black was used.
\end{enumerate}

\edit{Traditionally, researchers have characterized and modelled the polymer melts based on rheometric data from small amplitude oscillatory shear experiments (SAOS), which provide information about the linear viscoelastic behaviour of the materials.} \PoF{This rheological information has been proven to be useful in predicting the flow of monodisperse and bidisperse linear polymer melts; however, incorporating the nonlinear rheology is necessary to predict accurately the flow of polydisperse systems, such as those industrial melts, in extrusion flows~\cite{RobertsonJOR2017}.} Moreover, it is impossible to adequately describe the highly complex dependency on strain and strain rate experienced by a fluid element of viscoelastic material in a strong extensional flow using a shear rheological characterisation exclusively~\cite{GalindoRosales2013}. Thus, a complete rheological characterization consisting of a combination of steady shear flow curves and steady extensional flow experiments followed the protocols is \edit{detailed below, and is also} present in our previous works \cite{schuller_add_ma,sietsepaper}.

\edit{Viscosity curves under steady shear conditions were obtained at the working temperature (Table \ref{tab:temp_polymers}) for each polymer using an ARES G2 rotational rheometer. Parallel plates with a 12 mm diameter and a 1 mm gap were used. For shear rheological tests, specimens were shaped into disks (12 mm diameter, 1.5 mm thickness) through hot compression moulding. The specimens were pre-conditioned in a well-ventilated oven and then compression-molded, followed by cooling with a water circulation system.}

\edit{For the elongational viscosity measurements, a rotational rheometer (ARES G2) equiped with the TA Instruments Extensional Viscosity Fixture (EVF)~\cite{EVFpatent,EVFTAinstr1,EVFTAinstr2} was employed, imposing various strain rates (0.1, 0.3, 1, and 3 s$^{-1}$) to fresh samples. Specimens (8-10 mm width, 1 mm thickness, 18 mm length) were die-cut from compression-molded plates, pre-dried, and cooled with a water circulation system before testing.}

\begin{table}[ht!]
    \centering
    \caption{\edit{Polymer melt working temperatures.}}
    \begin{tabular}{ccccccc} \toprule
\textbf{Material} & PET-G  & PLA    & PC    & PA6/66 & ABS   & PET-CF \\\midrule

\textbf{Temp. [$^\circ$C]}     &       240   &    205 \cite{sietsepaper}    &   260 \cite{schuller_add_ma}   &   243    & 240   & 265\\\bottomrule
\end{tabular}
\label{tab:temp_polymers}
\end{table}

\edit{Among the plethora of constitutive models available in the literature for polymer melts, the Giesekus model was chosen} (Eq.~\ref{Eq:giesekus}), as in our previous works \cite{schuller_add_ma,sietsepaper}: 

     \begin{equation}
         \boldsymbol{\tau} + \lambda \overset{\triangledown}{\boldsymbol{\tau}}+\alpha\frac{\lambda}{\eta_0}\left(\boldsymbol{\tau}\cdot\boldsymbol{\tau}\right) =\eta_0 \left(\nabla\boldsymbol{u}+\nabla\boldsymbol{u}^T\right),
         \label{Eq:giesekus}
    \end{equation}
    
\noindent where $\eta_0$ is the zero-shear viscosity, and $\lambda$ is the relaxation time, both determined from the relaxation spectrum in steady-state; the dimensionless Giesekus-model mobility factor ($\alpha$) governs the extensional viscosity and the ratio of the second normal stress difference to the first. This constitutive model\edit{, already} incorporated into the \textit{constitutiveEquations} library of rheoTool~\cite{rheoTool}, can predict shear-thinning, shear first and second normal-stress differences\edit{~\cite{understanding}. According to \citep{Maja}, both the Giesekus and Rolie-Poly models~\cite{LIKHTMAN20031} achieve similar results; however, the Giesekus model is more attractive to carry out simulations because it needs fewer parameters. Moreover, we have already validated the potential of the Giesekus model to predict the growth of the vortex size produced by viscoelastic fluids flowing through planar contractions microfluidic channels (Figure 5 in \citep{schuller_add_ma}); additionally, the Giesekus model was also able to predict the extrudate-swell measurements for polycarbonate (Figure 10 in \cite{schuller_add_ma}). Furthermore, we have recently reported that the Giesekus model can reasonably predict the pressure inside the nozzle for PLA (Figure 8 in \citep{sietsepaper}), experimentally measured with a custom nozzle design. Based on this, we are confident that the Giesekus model captures the physics of the polymer melt flow through the nozzle in real FFF printing conditions.} 

\edit{As the printing process imposes a complex flow, combining extensional and shear flows, the Giesekus model has to be able to fit both kinds of non-linear rheometric datasets to provide meaningful predictions. Thus, the rheometric datasets from shear and extensional rheological characterisations determined the value of parameters of the Giesekus model and the number of modes. It is important to note that in its current version, \textit{rheoTool} does not perform the automatic fitting of the constitutive model parameters from the user-input experimental data. It is currently a hot research topic developing new artificial intelligence tools for fitting constitutive models' parameters to non-linear data sets, such as those coming from LAOS experiments~\cite{JohnAI}. Thus, no routine is available for providing the fitting parameters of a constitutive model based on nonlinear rheometric data combining shear and extensional information.}

\begin{figure}[hb!]
     \centering
         \includegraphics[width=0.8\textwidth]{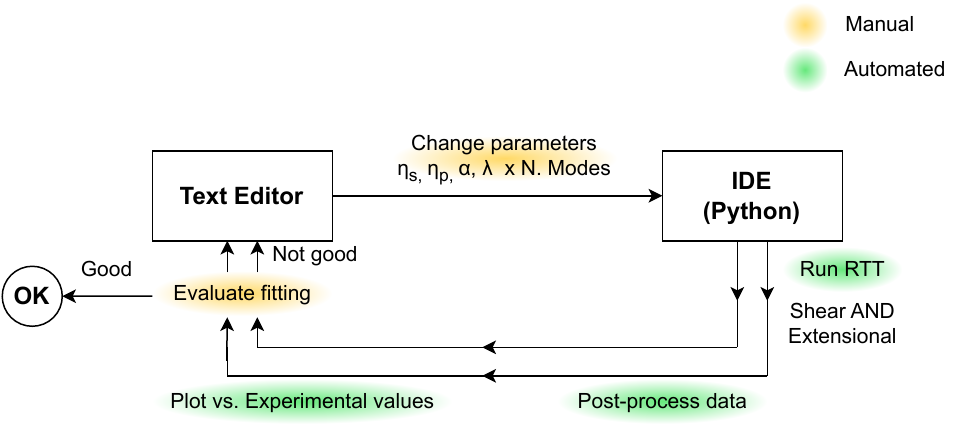}
        \caption{\edit{Work flow for determining the values of the Giesekus model parameters based on both shear and extensional rheometric data set.}}
        \label{fig:rtt_flowchart}
\end{figure}

\edit{Figure~\ref{fig:rtt_flowchart} shows the iterative process for determining the values of the Giesekus model, which combines the manual modification of the parameters values and the automatic production of the numerical prediction to the experimental datasets.} \textit{rheoTestFoam} was used to compare numerical responses with the rheometric datasets under the same experimental \edit{conditions.} The initial approach to fitting the models was fully manual, involving the user in all the steps, leading to a repetitive and tedious process. \edit{To improve the process}, a collection of Python scripts was developed to run and plot the data automatically (Figure~\ref{fig:rtt_flowchart}), allowing for a more visual iterative process drastically reducing the model creation time \cite{foamscripts}. \edit{However, the final evaluation of the numerical prediction was qualitative.}

\begin{table}[ht!]
    \centering
    \caption{Polymer melt parameters (All materials).}
    \begin{tabular}{ccccccc} \toprule
\textbf{Material} & PET-G  & PLA    & PC    & PA6/66 & ABS   & PET-CF \\\midrule
Model    & \multicolumn{6}{|c|}{Giesekus} \\
Num. Modes & 3      & 3      & 4     & 2      & 2     & 2      \\
Avg. $\eta$ & 600    & 2220   & 563   & 1300   & 43900 & 1038   \\
Avg. $\lambda$ & 0.355  & 0.209  & 0.713 & 0.518  & 13.27 & 27.21  \\
Avg. $\alpha$  & 0.0331 & 0.0133 & 0.233 & 0.239  & 0.107 & 0.018  \\ \bottomrule
\end{tabular}
\label{tab:prop_polymers}
\end{table}

\begin{figure}[htbp!]
\centering
\includegraphics[width=0.7\linewidth]{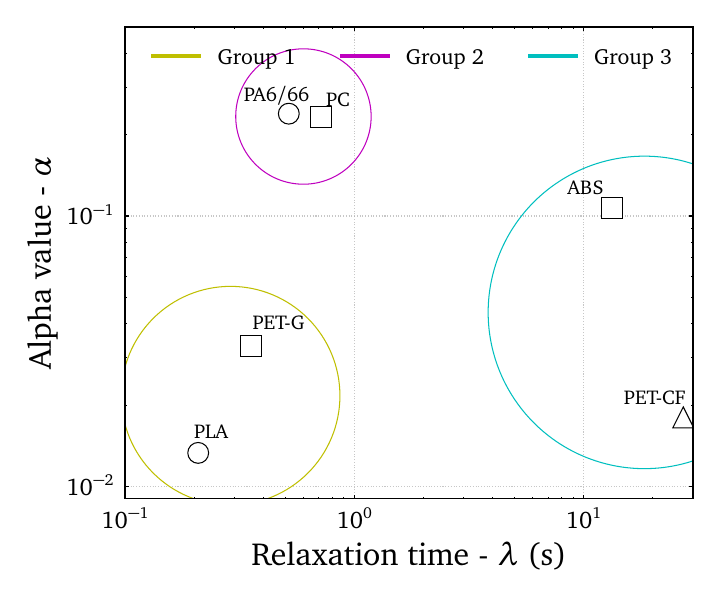}
\caption{\edit{Parametric map $\alpha-\lambda$, which allows for }grouping the materials by property proximity. In regards to crystallinity: Amorphous - Square, Semi-crystalline - Circle, Composite - Triangle.}
\label{fig:cluster}
\end{figure}

\begin{figure*}
    \centering
    \begin{tblr}{Q[l,h]Q[c,h]Q[c,h]}
        \textbf{Group 1} & \parbox{0.28\textwidth}{\includegraphics[width=0.28\textwidth]{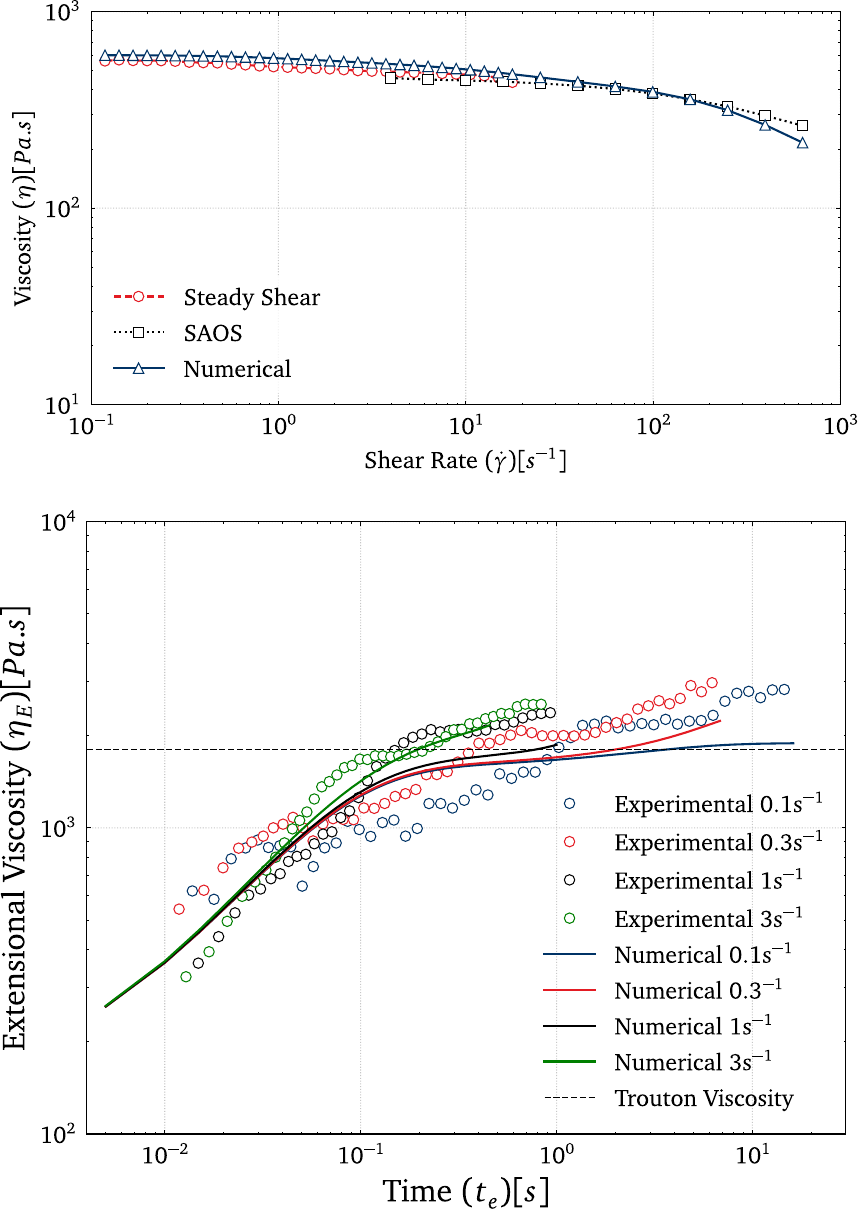}}& \parbox{0.28\textwidth}{\includegraphics[width=0.28\textwidth]{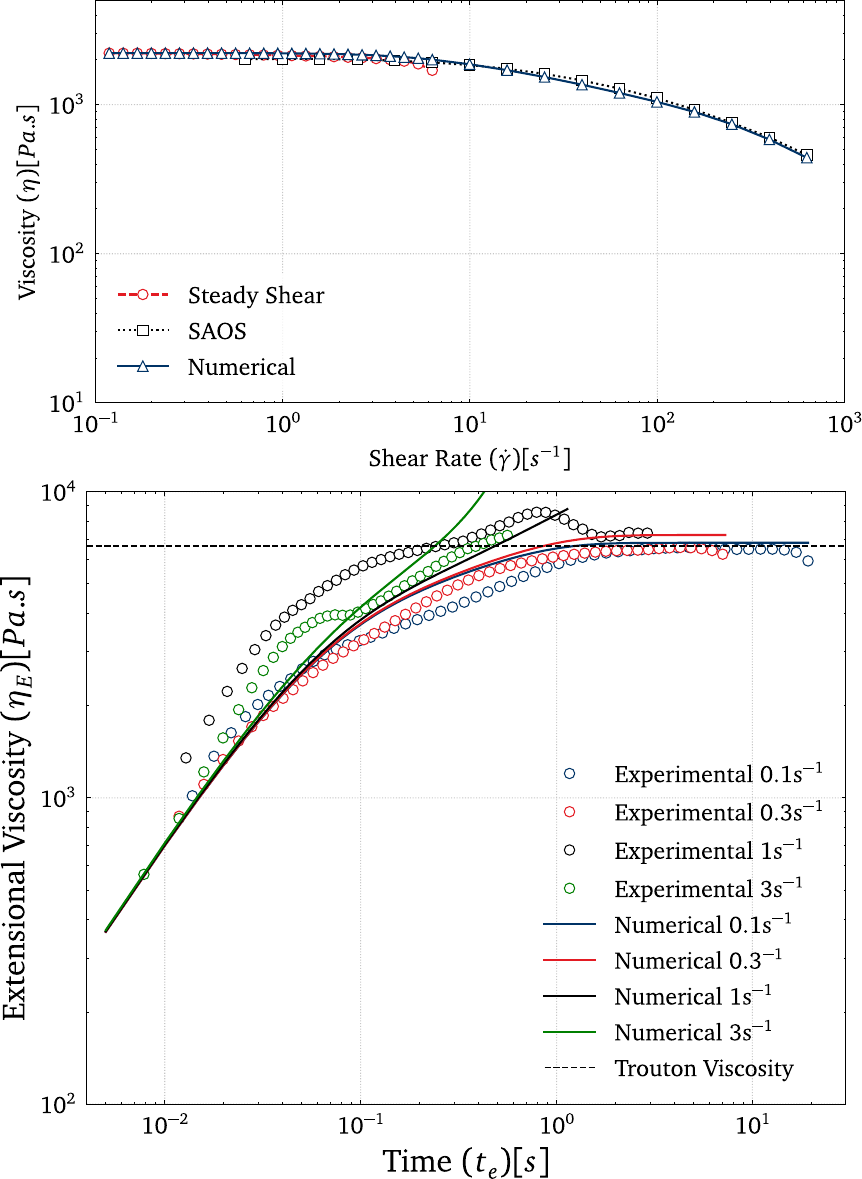}}\\
          & a) & b) \\
        \hline
        {\textbf{Group 2}}& \parbox{0.28\textwidth}{\includegraphics[width=0.28\textwidth]{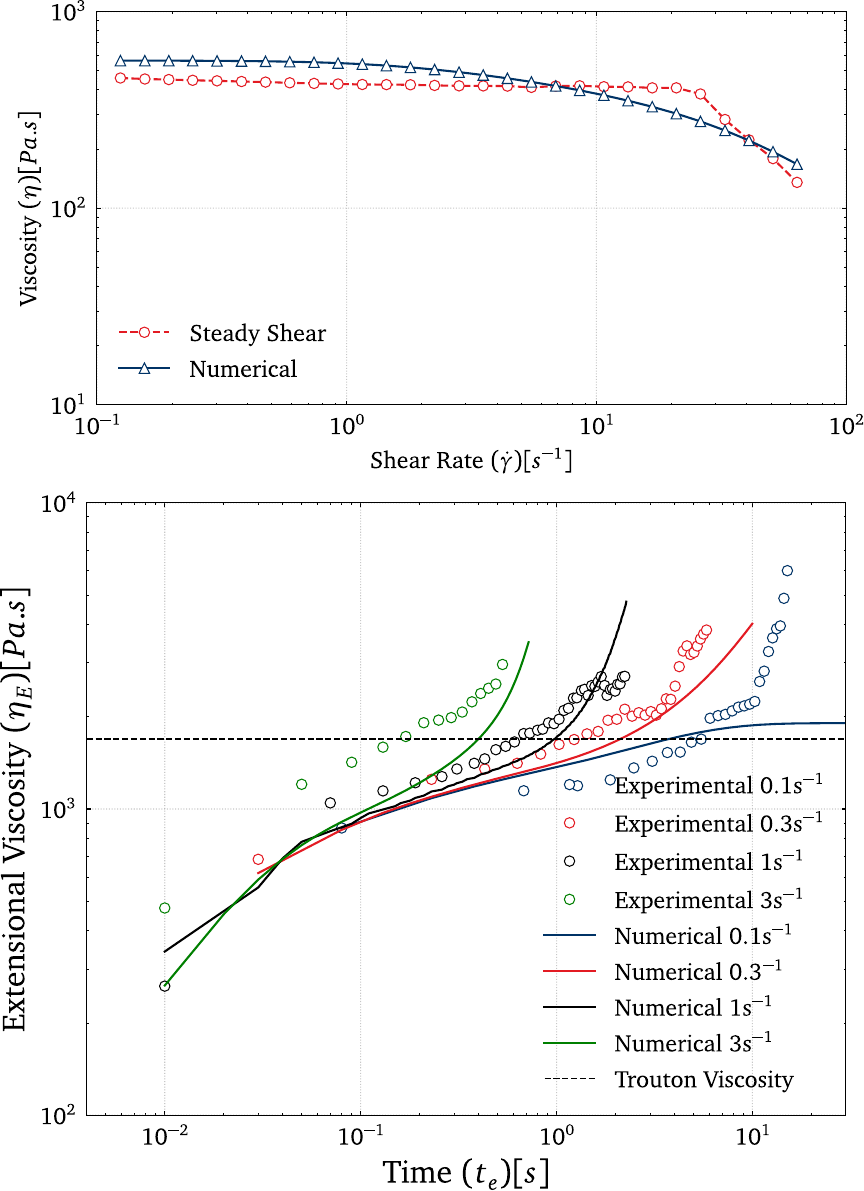}}& \parbox{0.28\textwidth}{\includegraphics[width=0.28\textwidth]{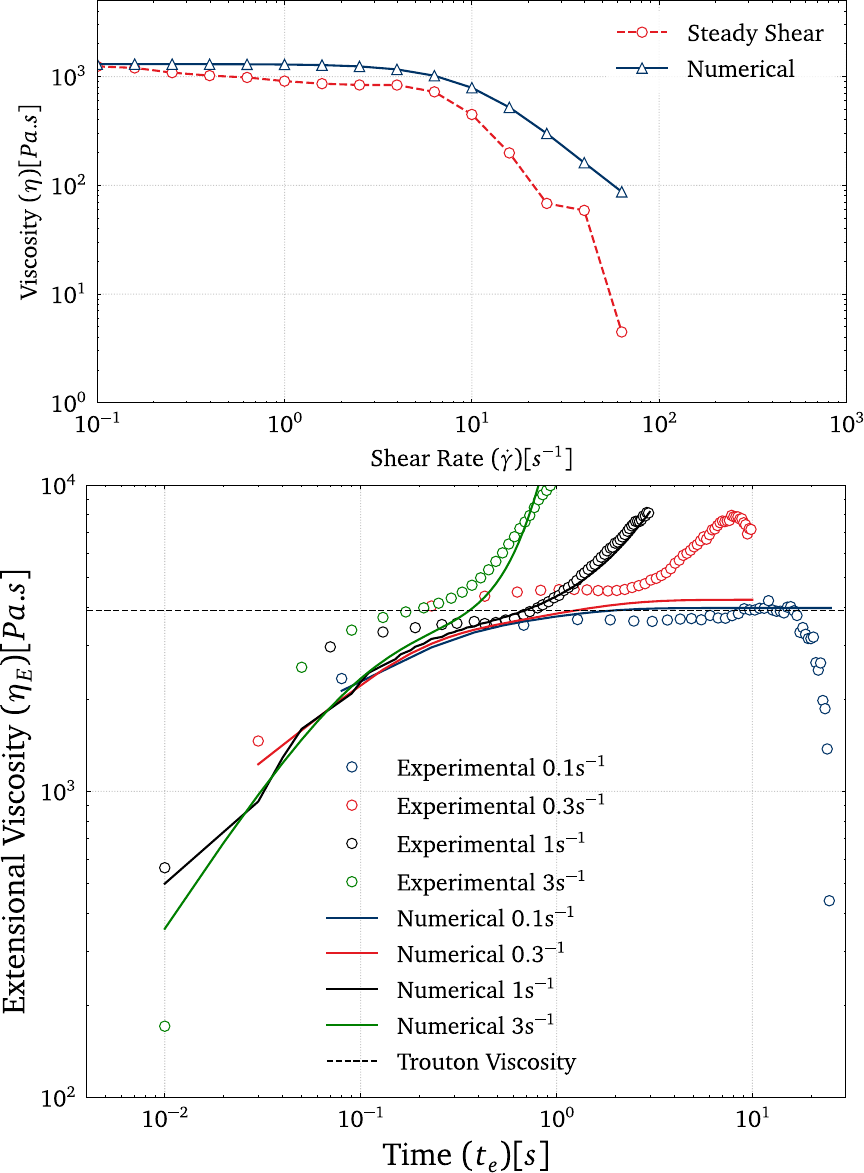}}\\
          & c)  & d) \\
        \hline
        {\textbf{Group 3}}& \parbox{0.28\textwidth}{\includegraphics[width=0.28\textwidth]{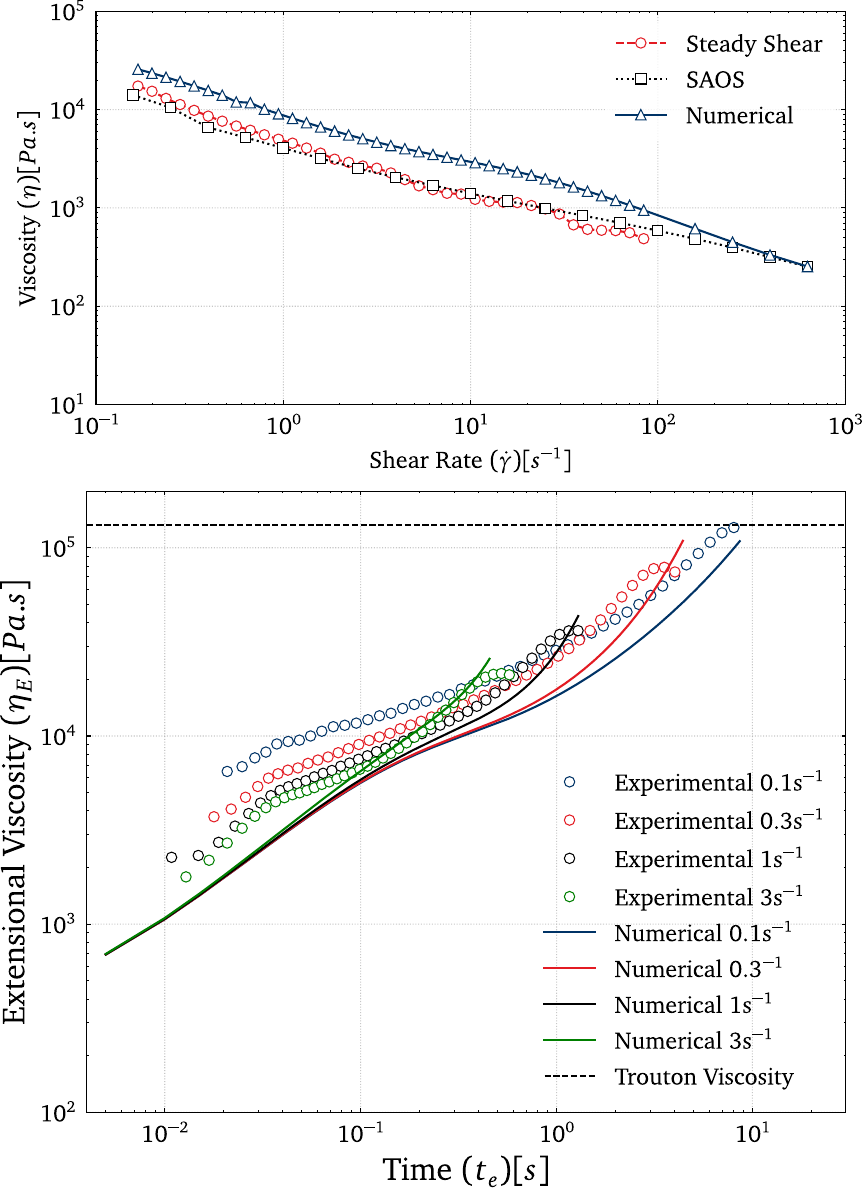}}& \parbox{0.28\textwidth}{\includegraphics[width=0.28\textwidth]{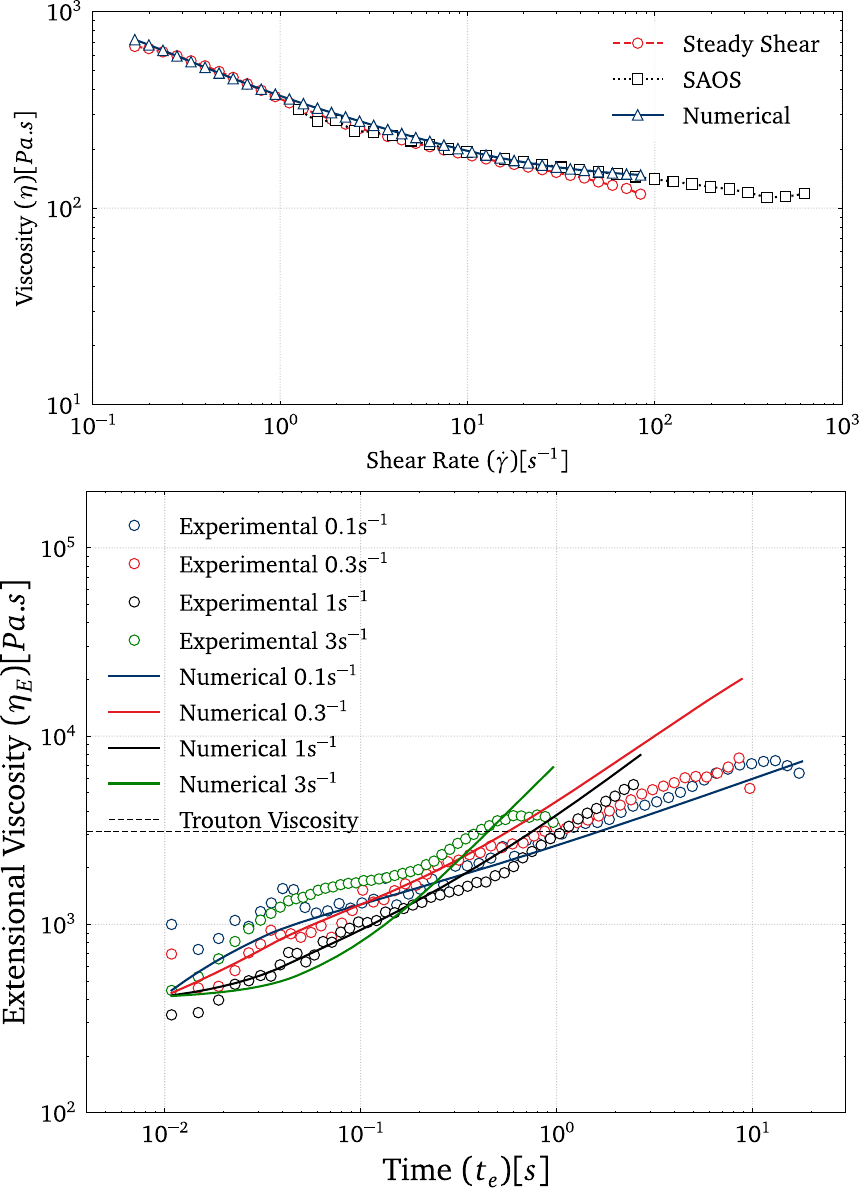}}\\
          & e)  & f) 
    \end{tblr}
    \caption{\edit{Steady shear and extensional viscosity curves for all materials. Experimental datasets (symbols) are compared against their numerical predictions (solid lines) obtained for the Giesekus model with the parameters given in Table~\ref{tab:prop_polymers}: a) PET-G, b)PLA, c) PC, d) PA6/66, e) ABS and f) PET-CF.}}
    \label{fig:visc} 
\end{figure*}

Table \ref{tab:prop_polymers} details the \edit{estimated values for model parameters for} each polymer\edit{, weight-averaged by $\eta$. Figure \ref{fig:cluster} presents the parametric map $\alpha-\lambda$, as a sort of Ashby diagram, a valuable tool for materials selection and design optimization~\cite{ASHBY201197}.} The goal is to group similar objects into clusters having similar values of the $\lambda$ and $\alpha$ parameters, enhancing our understanding of their rheological relationships. Figure \ref{fig:visc} shows \edit{a comparison between the experimental rheological characterization and the numerical predictions for each working polymer. The subplots of Figure \ref{fig:visc} were gathered based on} the groups formed in Figure \ref{fig:cluster}, confirming that similar locations in the $\alpha-\lambda$ parameters-map resulted in similar rheological behaviours.

\edit{Regarding the experimental extensional datasets, the reader must be aware of the fact that their quality is lower than the one from the shear viscosity curves due to difficulties associated with imposing purely extensional deformations in the EVF system,} \PoF{such as sagging due to gravitational effects. Having an excellent rheological dataset is paramount for fitting the parameters of the constitutive model in order to ensure the quality of the numerical predictions. The filament stretching approach~\cite{tirtaatmadja1993filament} has been proven to provide excellent results not only for Newtonian fluids and polymer solutions~\cite{anna2001interlaboratory}, but also for suspensions~\cite{Andrade2020}; Bach et al.~\cite{BachJOR2003} developed a prototype of filament stretching rheometer suitable for polymer melts at high temperatures, avoiding problems with temperature gradients, and end-plate slip and ensuring reliable measurement of deformation; unfortunately, this device is not currently commercialized.} 

Considering this, in the trial-error procedure for determining the model parameters for each polymer, the best possible approximation to the shear dataset was the first criterion for considering them reliable. Then, the trends defined by the extensional viscosity data were also tried to follow. Regarding the number of modes, we tried to keep them as low as possible, but we needed to increase them to improve the accuracy of the prediction. Deviations of the numerical predictions with the Giesekus model from the experimental measurements in \cite{schuller_add_ma, sietsepaper} are not associated with the quality of the approximation to the rheometric data, but either with the simplifications assumed in the numerical modelling of the physical problem, such as isothermal flow, and backflow issues.

\edit{Figure~\ref{fig:visc} also illustrates the transient evolution of extensional viscosity ($\eta^{+}_{E}\left(t,\dot\epsilon\right)$)~\cite{nomenclatureJOR} at various extension rates. 
For the materials in Group 1, it can be observed that all the curves overlap; additionally, the extensional viscosity reaches a steady state pointing towards a value respecting the Trouton ratio, i.e. $\mathit{Tr}=\frac{\eta_E^*}{\eta_0}$)~\cite{Trouton,Wisniak2001}, which can be considered as a sign of the measurement quality and that the polymer chains are excited within the linear viscoelastic region~\cite{Dealy1999, Claus2003}. Materials included in Group 2 and 3 exhibited a marked strain-hardening, exhibiting Trouton ratios larger than 3~\cite{Munstedt2023}.}

\FloatBarrier
\subsection{Nozzle flow simulations}

The nozzle geometry consists of an axisymmetric contraction as described in \cite{schuller_add_ma}. The tube length was adjusted to guarantee flow development at the start of the contraction region before the tapered section. Gravity was aligned with the nozzle's symmetry axis, and the cylindrical coordinate system's origin coincided with the beginning of the upstream length. Two boundary conditions were applied at the internal wall of the nozzle: a non-slip condition and zero pressure gradient. \edit{The non-slip condition means that the velocity of the polymer melt is the same as that of the solid wall, which in this case is zero, as the liquefier and the nozzle are stationary. Zero pressure gradient means no net change in pressure along the flow direction within the boundary layer; in other words, the pressure remains relatively constant as the fluid moves along the surface. In the context of the boundary layer, a zero pressure gradient condition implies that the pressure forces acting on the fluid in the direction parallel to the surface are relatively balanced, and there is no strong driving force due to pressure differences. The zero pressure gradient boundary layer is of interest in various engineering applications, and its study helps understand the behaviour of fluid flow near surfaces}. Additionally, at the inlet face, a uniform velocity profile ($\boldsymbol{V_{in}}=\overline{V_{in}}\boldsymbol{e_z}$) was enforced, with $\overline{V_{in}}$ representing the average velocity linked to the extrusion velocity $\overline{V_{ext}}$:


\begin{equation}
    \overline{V_{in}}=\overline{V_{ext}}\frac{D_c^2}{D_u^2}.
\end{equation} 

\noindent In this research, we examined nine different extrusion velocities spanning four orders of magnitude: $\overline{V_{ext}}$=\{0.01, 0.1, 1, 2, 5, 10, 30, 70, 110\} mm/s. The Reynolds and Weissenberg numbers corresponding to each velocity and material are detailed in \edit{Tables~\ref{tab:re0} and \ref{tab:wi0} (\ref{app:1})}. Additionally, we applied a zero gradient for pressure at the inlet. At the die exit, where the extruded material encounters atmospheric conditions, we set the pressure to zero, while the boundary condition for velocity was treated as outflow with a zero gradient.

The axisymmetric geometry allowed for efficient 2D numerical simulations instead of 3D, significantly reducing computational time. Consequently, the numerical 2D setup consisted of a narrow-angle wedge, one cell thick, running along the centerline. Wedge-type patches were used for both velocity and pressure, as previously validated in \cite{schuller_add_ma}. Similar to our prior studies~\cite{schuller_add_ma,sietsepaper}, a structured two-dimensional mesh was created using the \textit{blockMesh} utility, comprising three zones: 1) the straight upstream section, 2) the tapered area, and 3) the die region. A stream-wise stretch ratio was applied to enhance resolution at the die and abrupt contraction zone. A grid analysis ensured spatial convergence for the presented results in this work.


\edit{Isothermal and steady flow through the extrusion nozzle were assumed for all materials and simulations. Consequently, the energy equation was decoupled from the mass and momentum conservation equations. Treating the molten polymer as an incompressible fluid, the mass conservation equation was simplified as depicted in Equation \ref{eq: Continuity}:}
 
 \begin{equation}
  \label{eq: Continuity}
  \nabla \cdot \boldsymbol{u} =0.
\end{equation} 

\noindent \edit{Equation \ref{eq: Momentum} gives the momentum conservation equation:}
 
\begin{equation}
  \label{eq: Momentum}
  \rho \boldsymbol{u}\cdot\nabla \boldsymbol{u} =-\nabla P -\nabla\cdot\boldsymbol{\tau}.
\end{equation} 

\noindent \edit{In this context, we considered a steady-state flow ($\frac{\partial\boldsymbol{u}}{\partial t}=0$), while excluding the negligible effects of gravity. Here, we used the notation $P$ for pressure, $\rho$ for density, and the extra stress tensor $\boldsymbol{\tau}$, which encompasses contributions arising from the fluid deformation as defined in the Giesekus model found in Eq. \ref{Eq:giesekus}. }

These equations were solved using the rheoTool library~\cite{rheoTool,pimenta1} integrated with OpenFOAM\R \cite{openfoam}. A high-resolution scheme (CUBISTA) and the log-conformation formulation of the constitutive equation ensured numerical stability~\cite{AlvesetalAnnRevFluMech2021}.

\FloatBarrier
\section{Results and discussion}

The kinematics of the flow through the nozzle geometry is complex, i.e. at the wall, the polymer melt will undergo pure shear flow ($\dot\gamma=\frac{\partial u_z}{\partial r}$) and, at the centerline, the rate of deformation is purely extensional ($\dot\epsilon=\frac{\partial u_z}{\partial z}$). The complexity of the contraction flow is well represented by the flow-type parameter~\cite{ORTEGACASANOVA2019102}, defined as:

\begin{equation}
    \xi=\frac{\|\boldsymbol{D}\| - \|\boldsymbol{\Omega}\|}{\|\boldsymbol{D}\| + \|\boldsymbol{\Omega}\|},
\label{eq:flowtype}
\end{equation}

\noindent In this context, $\|\boldsymbol{D}\|=\sqrt{\frac{\boldsymbol{D}:\boldsymbol{D}^T}{2}}$ represents the magnitude of the rate-of-deformation tensor $\boldsymbol{D}=\frac{1}{2}\left(\nabla\boldsymbol{u}+\boldsymbol{u}^T\right)$, and $\|\boldsymbol{\Omega}\|=\sqrt{\frac{\boldsymbol{\Omega}:\boldsymbol{\Omega}^T}{2}}$ signifies the magnitude of the vorticity tensor $\boldsymbol{\Omega}=\frac{1}{2}\left(\nabla\boldsymbol{u}-\boldsymbol{u}^T\right)$, where $\boldsymbol{u}$ denotes the velocity field. Irrespective of the fluid's rheological characteristics and the nozzle's geometry, reducing the diameter induces a complex flow pattern. This pattern encompasses regions characterized by purely simple shear ($\xi=0$), areas of purely elongational flow ($\xi=1$), sections approaching solid-body rotation ($\xi=-1$), and regions displaying a combination of these behaviours \cite{schuller_add_ma,THOMPSON1999375,MOMPEAN2003151,ma15072580}. The contraction within the nozzle generates a diverse range of flow types, spanning from simple shear to extensional flow, encompassing the entire nozzle's cross-section.
That kinematics complexity results in a complex distribution of normal stress differences ($\Delta \tau=\tau_{zz}-\tau_{rr}$), that is, at the wall, the normal stress difference corresponds to the shear-induced ones ($\Delta \tau=N_1$); at the centerline, the normal stress difference is related to the extension of the molecules ($\Delta \tau=\sigma_E$); whereas at any location in between they are coupled \edit{\cite{Tseng2023}}.

\begin{figure*}
   \centering
\begin{tblr}{Q[l,b] Q[l,m] Q[l,m]}
\rotatebox{90}{\textbf{Group 1}}&
\includegraphics[width=0.425\textwidth]{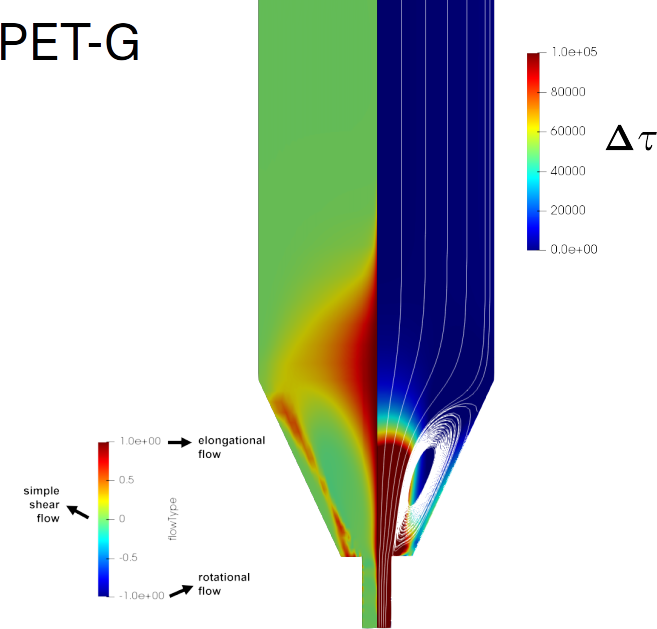}&
\includegraphics[width=0.44\textwidth]{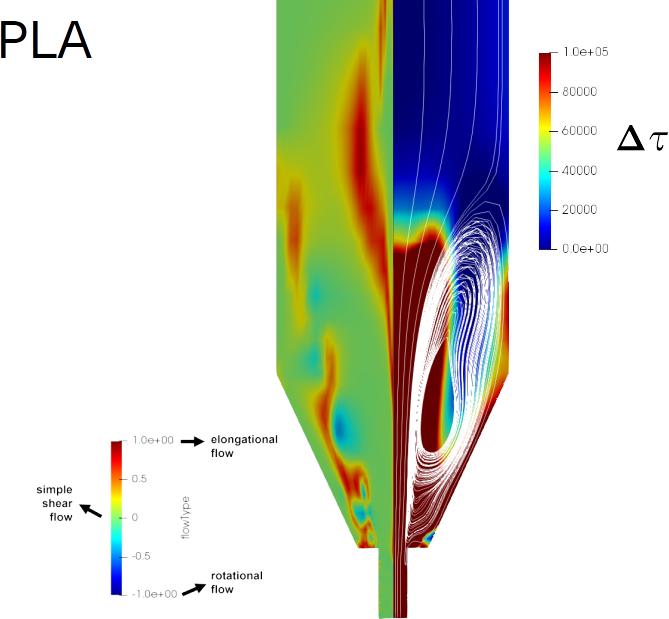}\\\hline
\rotatebox{90}{\textbf{Group 2}}&
\includegraphics[width=0.43\textwidth]{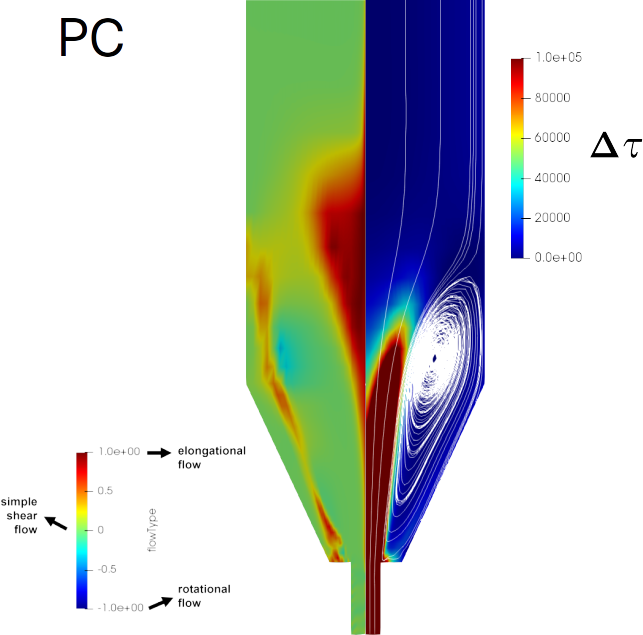}&
\includegraphics[width=0.425\textwidth]{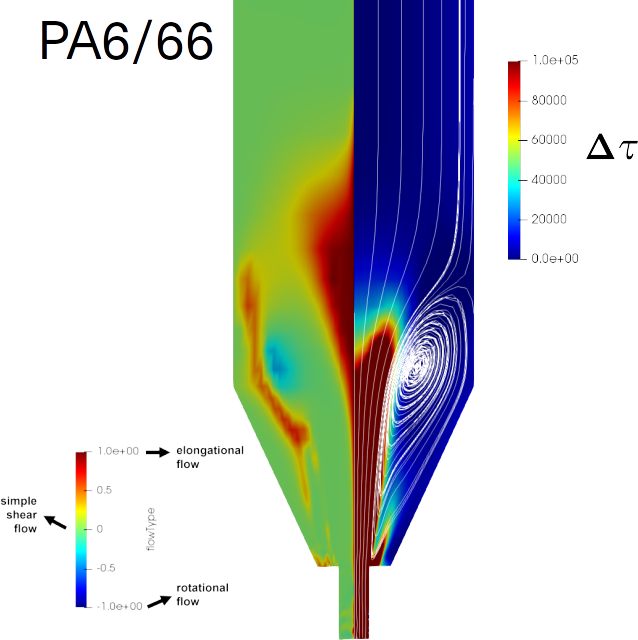}\\\hline
\rotatebox{90}{\textbf{Group 3}}&
\includegraphics[width=0.46\textwidth]{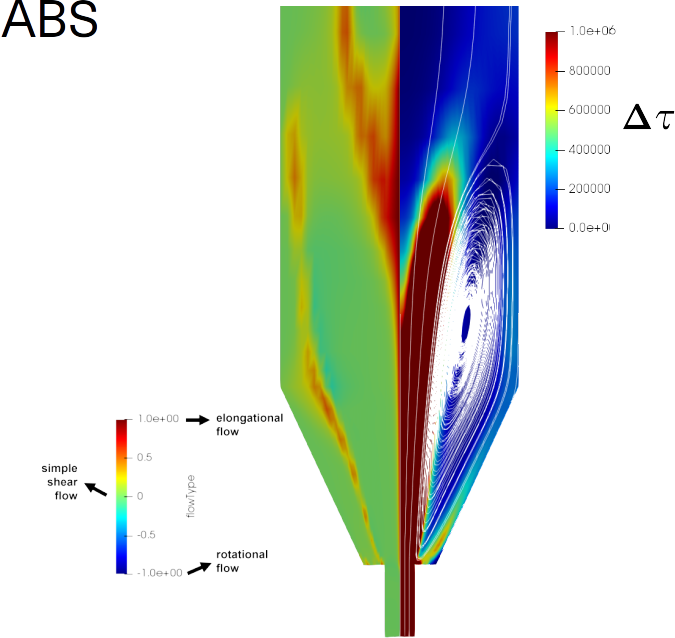}&
\includegraphics[width=0.425\textwidth]{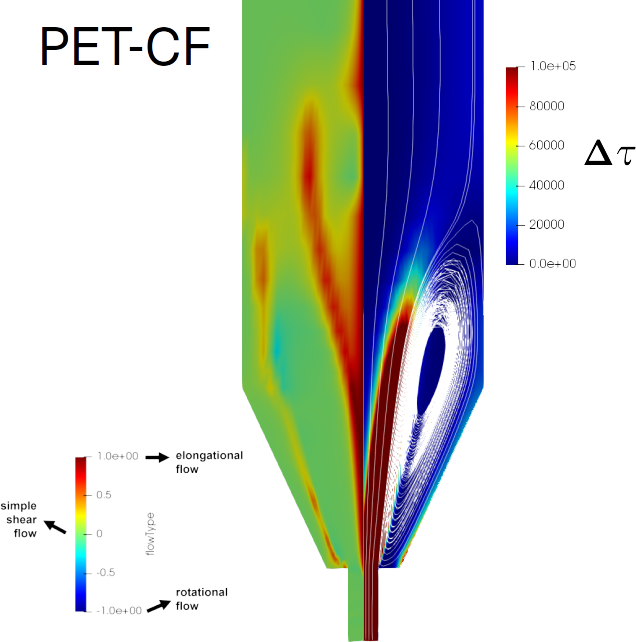}
\end{tblr}
    \caption{\edit{Flow-type (left \edit{half}), $\tau_{zz}$- $\tau_{rr}$ and streamline (right \edit{half}, layered) surface plots (110mm/s) for all materials considered in this study.}}
    \label{fig:flowtype} 
\end{figure*}

Figure~\ref{fig:flowtype} shows a composition of the contour plots corresponding to the gradient of flow type (left half) and the normal stress differences (right half) for all the materials considered in this study. Different patterns in the flow type can be observed depending on the material extruded; as expected, the flow type is mainly dominated by the shear flow (green), although important portions of the fluid domain, axisymmetric and with a dart shape, are pure extensional flow (red). As mentioned above, that complexity leads to a complex rate of deformation distribution, which, depending on the rheological properties of the fluid, results in a complex distribution of stress differences, depending on the material's rheological properties. 

\noindent It can be observed in \edit{Figure~\ref{fig:visc}} that PLA and PET-G (Group 1) exhibit shear thinning\edit{,} and a soft-curved and asymptotic increase of the extensional viscosity, which resulted in a squared normal stress difference distribution centred at the axis of symmetry \edit{(Figure~\ref{fig:flowtype})}. PC and PA6/66 (Group 2) also exhibited shear thinning under shear flow but strain hardening under extensional flow \edit{(Figure~\ref{fig:visc})}; this rheological behaviour resulted in an indent at the centerline in the previous squared normal stress difference, compared to that shown by PLA and PET-G \edit{(Figure~\ref{fig:flowtype})}. \edit{Finally,} ABS and PET-CF (Group 3) presented shear thinning, but a steep increase of the extensional viscosity with the accumulated deformation \edit{(Figure~\ref{fig:visc}), which } resulted in an even more pronounced indentation in the normal stress difference contour plot at the centerline \edit{(Figure~\ref{fig:flowtype})}. 

\noindent \edit{In the elongational characterization of the polymer melts, the extensional viscosity is plotted against the accumulated deformation ($\epsilon$). The accumulated deformation result from integrating the extension rate ($\dot\epsilon$) over the time of the experiment:}
\begin{equation}
    \edit{\epsilon=\int_{0}^{t}\dot\epsilon dt,}
\end{equation}
\noindent \edit{as in the rheometric experiment the extension rate is kept constant, then $\epsilon=\dot\epsilon t$. This relationship is essential in understanding how the material accumulates deformation in an elongational flow field, providing insights into the total extension experienced by the material over a given period. It is particularly relevant in polymer processing, fluid dynamics, and other fields where the deformation of materials is a critical aspect of the study.} 
\noindent \edit{In the centerline of the nozzle geometry, the fluid undergoes a shear-free extensional flow, in which the extension rate depends on the position in the tapered region~\cite{schuller_add_ma}. The average extension rate ($\overline{\dot\epsilon}$) increases with the volumetric flow rate ($Q$), according to the following equation:}
\begin{equation}
    \edit{\overline{\dot\epsilon}\approx \frac{4Q}{\pi L}\left(\frac{1}{D_{c}^{2}}- \frac{1}{D_{u}^{2}}\right).}
\end{equation}
\noindent \edit{Thus, on average, increasing the flow rate would increase the accumulated deformation for a constant residence time ($\overline{\epsilon}=\overline{\dot\epsilon}~ t_{res}$). The residence time ($t_{res}$) of the polymer melt flowing through the nozzle can be calculated by dividing the volume of the nozzle ($V$) by the flow rate. The volume of the nozzle is the same for all the cases and all the materials. Thus, the residence time is inversely proportional to the volumetric flow rate ($t_{res}=V/Q$). This means the residence time decreases as the volumetric flow rate increases, and vice versa. Therefore, the average accumulated deformation in the nozzle ($\overline{\epsilon}$) does not depend on the flow rate but the geometric dimensions of the nozzle:}
\begin{equation}
    \edit{\overline{\epsilon}\approx \frac{4V}{\pi L}\left(\frac{1}{D_{c}^{2}}- \frac{1}{D_{u}^{2}}\right).}
\end{equation}
\noindent \edit{It has been demonstrated that the averaged accumulated deformation remains independent of the flow rate for a specified nozzle shape.} 

\noindent \edit{Based on the arguments presented above, this result can be directly correlated with the extensional rheometric information as follow: a material with a steeper extensional viscosity curve will provide a larger resistance to flow under extensional flow for a given accumulated deformation, that is the same nozzle shape.}\\


\noindent Due to the viscoelastic nature of the polymers, upon a critical flow rate, upstream vortex structures were observed for all of them (Figure~\ref{fig:flowtype}). The tip of the normal stress difference distribution marked the tip of the upstream vortex structure, confirming that the presence of upstream vortices in viscoelastic contraction flows is due to elastic instabilities. Further increasing the flow rate leads to an unsteady motion of the vortex and oscillating values in the pressure drop, as previously reported~\cite{rothstein,sietsepaper}.

\begin{figure}[ht!]
\centering
\includegraphics[width=0.65\linewidth]{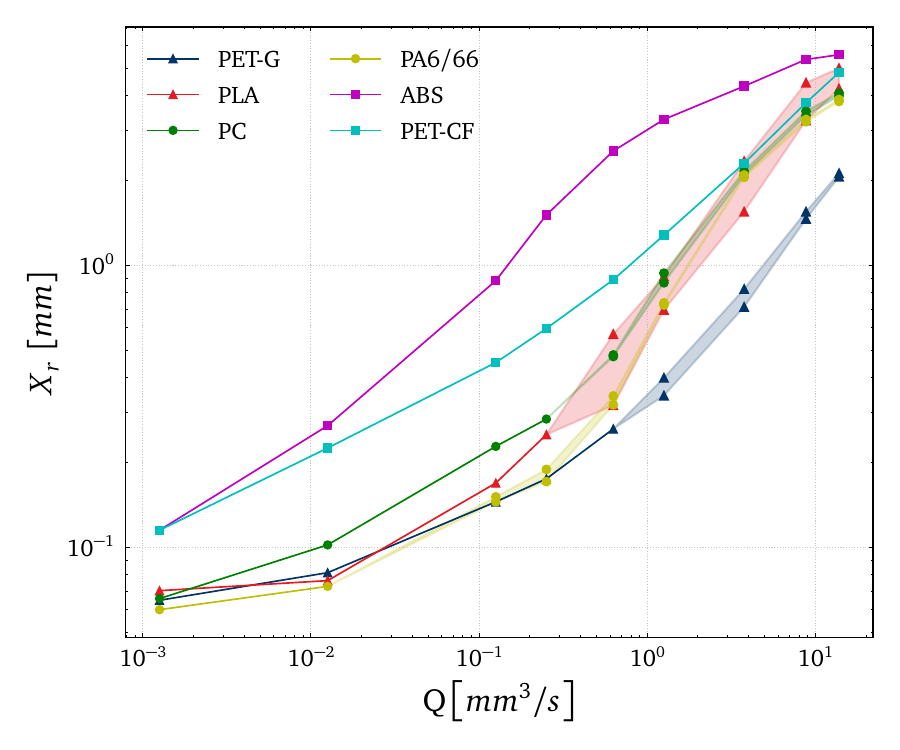}
\caption{Numerical upstream vortex lengths for the different materials.\\ Group 1 - triangles; Group 2 - circles; Group 3 - squares.}
\label{fig:vortex}
\end{figure}

Figure \ref{fig:vortex} shows the typical \textit{S-shaped} curve for the upstream vortex size ($X_{r}$~[mm]) with a growing flow rate (Q~[mm$^3$/s]) for all materials, which tends asymptotically to the Moffatt vortex size at low flow rates and to a saturation size at high flow rates~\cite{oliveira1}. The combination of considerably high relaxation time and viscosity resulted in a faster growth rate of the upstream vortex size, as could be observed for Group 3 of polymers, whereas relatively low viscosities and relaxation time exhibited a lower rate of growth in upstream vortex structures (\textit{e.g.} PET-G). The sequence observed in Figure~\ref{fig:vortex} can be inferred from the product $\eta \cdot \lambda$ (Table~\ref{tab:prop_polymers}), which is an estimation of the zero-shear first normal stress coefficient ($\Psi_{10}$~\cite{nomenclatureJOR}, ~\ref{app:1}); thus, the larger the first normal stress coefficient is, the more prone is the polymer to develop upstream vortices~\cite{rothstein1}.

\FloatBarrier

In FFF, the solid thermoplastic filament is melted in the liquefier and extruded through a nozzle. The rollers push down the solid filament, which works as a piston to impose a pressure-driven flow to the polymer melt. It has been previously reported~\cite{sietsepaper} that the total pressure drop measured in the feeders can be decoupled as a sum of the pressure drop in the liquefier and the pressure drop in the nozzle, being the latter very close to the one predicted by the numerical model proposed here. In a cylindrical coordinate system, the conservation momentum equation for the steady-state Poiseuille flow of viscoelastic material through the nozzle gets reduced to the following ones, corresponding to $r$ and $z$-directions, respectively~\cite{understanding}:

\begin{equation}
\rho\left(u_r\frac{\partial u_r}{\partial r}+u_z\frac{\partial u_r}{\partial z}\right)\approx-\frac{\partial P}{\partial r}-\frac{1}{r}\frac{\partial N_2}{\partial r}-\frac{\partial\tau_{zr}}{\partial z}
\label{eq:Momentum_poiseuille_r}
\end{equation}

\begin{equation}
\rho\left(u_r\frac{\partial u_z}{\partial r}+u_z\frac{\partial u_z}{\partial z}\right)\approx-\frac{\partial P}{\partial r}-\frac{1}{r}\frac{\partial }{\partial r}\left(r\tau_{rz}\right)-\frac{\partial\tau_{rr}}{\partial z}
\label{eq:Momentum_poiseuille_z},
\end{equation}

\noindent where $u_r$ and $u_z$ must satisfy the mass conservation equation:

\begin{equation}
\frac{\partial u_r}{\partial r}+\frac{\partial u_z}{\partial z}=0.
\label{eq:Mass_poiseuille}
\end{equation}

\noindent The shear viscosity ($\eta\propto\tau_{rz}$) and the first normal stress difference ($\Delta \tau=\tau_{zz}-\tau_{rr}\approx\tau_{zz}$) are responsible for the pressure gradient in the direction of the flow (Eq.~\ref{eq:Momentum_poiseuille_z}), being $\Delta \tau$ dependent on the kind of deformation exerted to the polymeric molecules, i.e. extensional at the centerline, pure shear at the wall, or a combination anywhere in between. Therefore, the larger the shear viscosity and the larger the area occupied by large values of first normal stress differences (Figure~\ref{fig:flowtype}), the more significant the pressure drop will be.

\begin{figure}[ht!]

\centering
\begin{subfigure}{.5\textwidth}
    \centering
    \includegraphics[width=\textwidth]{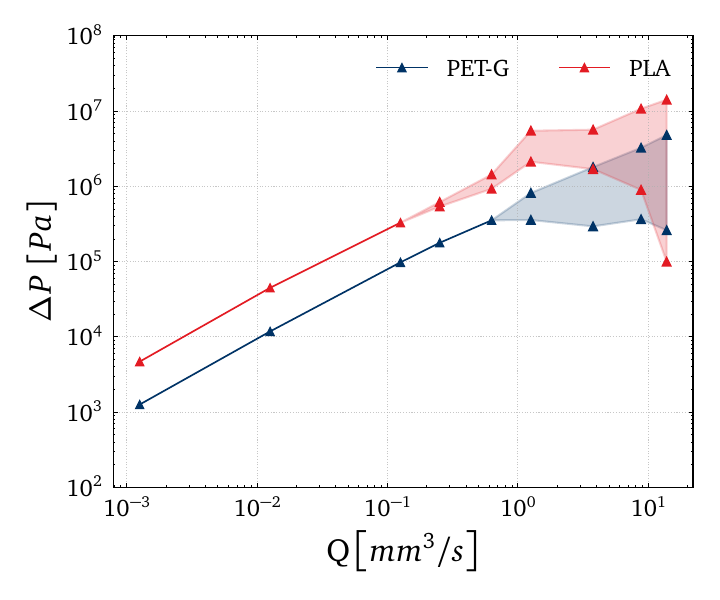}
    \caption[short]{}
\end{subfigure}%
\begin{subfigure}{.5\textwidth}
    \centering
    \includegraphics[width=\textwidth]{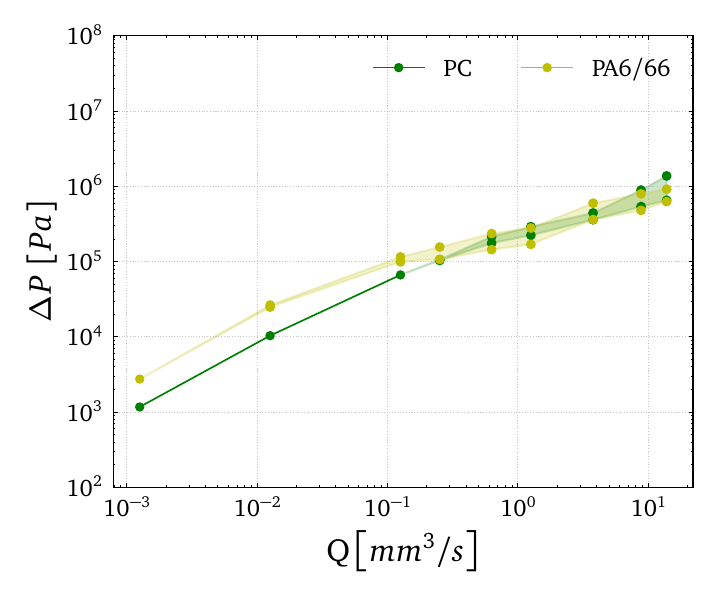}
    \caption[short]{}
\end{subfigure}\\ \vspace{1em}
\begin{subfigure}{.5\textwidth}
    \centering
    \includegraphics[width=\textwidth]{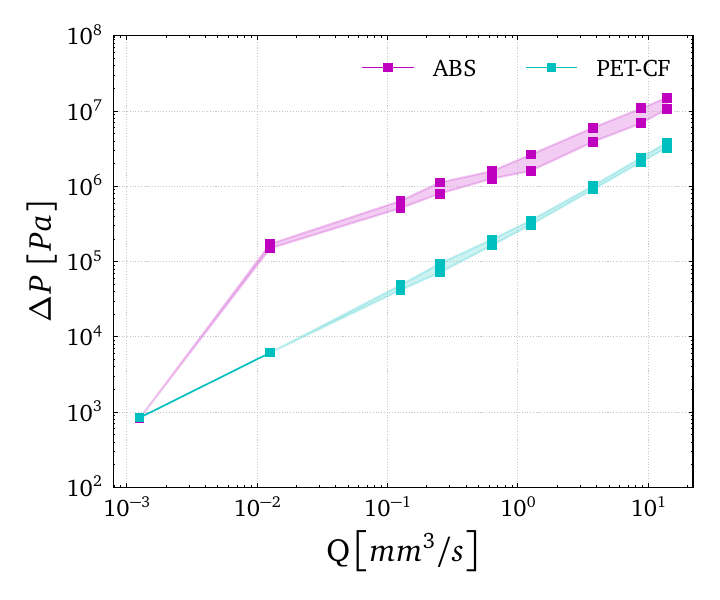}
    \caption[short]{}
\end{subfigure}%
\caption[short]{Pressure drop against flow rate for all the polymers: (a) Group 1 - Plug shape in $\tau_{zz}$- $\tau_{rr}$ plots; (b) group 2 - Short arrow shape in $\tau_{zz}$- $\tau_{rr}$ plots; (c) Group 3 - Long arrow shape in $\tau_{zz}$- $\tau_{rr}$ plots.} 
\label{fig:pressure_ranges}
\end{figure}

\noindent Figure~\ref{fig:pressure_ranges} shows the dependence of the pressure drop with the imposed flow rate, and it confirms the previous analysis, showing that ABS, PLA and PET-CF exhibit more significant pressure drops, followed by PET-G, PC and PA6/66. It is important to note that Figure~\ref{fig:pressure_ranges} accounts for the pressure drop within the tapered region and the pressure drop in the straight parts of the nozzle. The shear thinning behaviour present in all the polymeric materials provokes a decrease in pressure drop due to viscous contribution; upon a critical flow rate, the elastic instabilities are triggered, and the excess pressure drop is activated~\cite{McKinley1991,Alves2007}. Further increasing the flow rate leads to an unsteady motion of the vortex and, subsequently, oscillating values in the pressure drop~\cite{sietsepaper}, which can be associated with elastic turbulence~\cite{Groisman2000-ym,Larson2000-gm}. A low value in the $\alpha$ parameter implies a less shear thinning in viscosity, and the fluid tends to a Boger-fluid-like behaviour; thus, the material would be more prone to trigger elastic instabilities and turbulence~\cite{PENG2021104571}.

\edit{By combining the data presented in Figure~ \ref{fig:pressure_ranges} with the viscosity curves depicted in Figure~\ref{fig:visc}, one can observe} that the larger the viscosity of the polymer, the larger the pressure drop. A stronger shear thinning behaviour (higher values in the $\alpha$ parameter) results in a bend of the $\Delta P-Q$ curve, whereas low values of $\alpha$ lead to elastic turbulence, with oscillating and larger pressure drops. This, together with the shapes of normal stress difference plots, allows for the understanding that a more squared/plug-shaped region in the center of the tapered region, along with the approximation of the vortex to the center of the channel (more pronounced in the PLA case (Figure \ref{fig:flowtype})), leads to a higher magnitude of elastic turbulence/instabilities in the reentrant corner, a critical region of the flow.







\begin{figure}[ht!]
\centering
\includegraphics[width=0.8\linewidth]{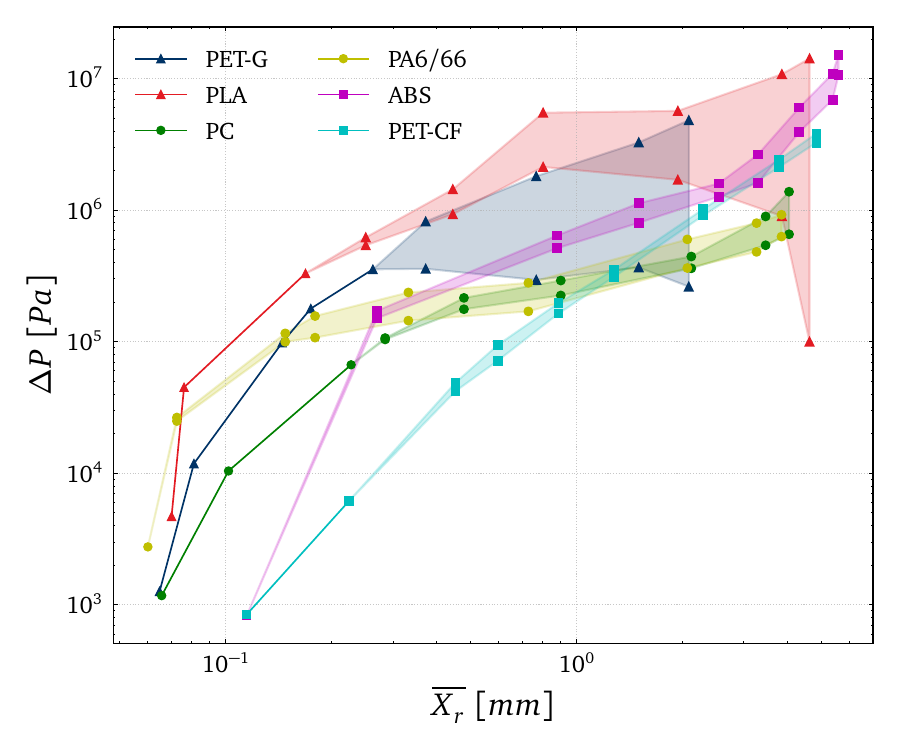}
\caption{Pressure \textit{vs} vortex size, for all materials.\\ Group 1 - triangles; Group 2 - circles; Group 3 - squares.}
\label{fig:vortex_vs_pressure}
\end{figure}

\edit{By plotting the vortex size} against the pressure drop for all materials \edit{(Figure \ref{fig:vortex_vs_pressure})}, it is possible to attest that polymers with relatively low relaxation time ($\lambda<1$~s) are more sensitive, and the pressure drop increases drastically for a small increase in vortex size in its initial stages. This increase then becomes more gradual as the recirculation grows. 

\edit{The shaded regions in Figures~\ref{fig:vortex},\ref{fig:pressure_ranges} and \ref{fig:vortex_vs_pressure} represent the printing conditions when elastic turbulence is dominating. In that region, the pressure drop values and the vortices for this nozzle begin to oscillate and the shaded regions indicate the space between the minimum and maximum values recorded in the simulations, as in \cite{sietsepaper}.}

Although die-swell was not considered a matter of investigation in this work, it can be an important parameter in FFF, particularly on the fiber orientation and the resulting mechanical properties~\cite{HELLER2016252}. Die-swell is not affected by the presence of the upstream vortices~\cite{james2021}, and it has been reported in the literature~\cite{TOME2019104} that the swelling ratio decreases as the value of the $\alpha$ parameter increases. Therefore, according to Table~\ref{tab:prop_polymers}, the larger die-swell would be expected for polymers belonging to Group 2, followed by ABS, PET-G, PET-CF and, finally, PLA.

\section{Conclusions}


This study has conducted a thorough numerical examination of upstream vortex formation in a standard Fused Filament Fabrication (FFF) nozzle during stable flow conditions. The investigation utilized the Giesekus constitutive model, incorporating data from steady simple shear experiments and transient extensional viscosity experiments conducted at different extension rates.

Significant contributions have been made to the body of knowledge previously derived from prior work \cite{schuller_add_ma,rooooooooy}. It has been demonstrated that even under steady flow conditions at extrusion velocities representative of practical scenarios, growing upstream vortices are influenced by elastic instabilities. These instabilities manifest as physical disturbances in pressure, leading to the phenomenon known as ``elastic turbulence''.

Complex flow patterns induced within the nozzle, ranging from simple shear to extensional flow, have been critically examined. These flow phenomena affect the behaviour of molten polymer from the nozzle walls to the axis (as depicted in Figure \ref{fig:flowtype}). This intricate behaviour persists through the tapered region and further upstream into the liquefier, while the flow inside the die is primarily characterized by simple shear.

Two key observations have arisen from the analysis of flow types and variations in normal stress: Firstly, the vortex tip marks the culmination of elastic stresses induced by extensional flow, and secondly, the shape of the stress contour plot has a direct impact on the proximity of the upstream vortex to the nozzle centerline. These observations are further supported by the consistent appearance of rheological groups identified during the characterization. In the examination of the plots displaying normal stress surfaces, the intricate interplay between these stresses surrounding the contraction has been discerned. This relationship sheds light on the rheological properties of the fluid and contributes significantly to variations in pressure drop and the emergence of instabilities. This fluid characteristic appears to correlate with elevated excess pressure drop, potentially causing operational challenges during printing. 

Furthermore, the sensitivity of various materials to the creation of recirculation vortices has been uncovered. A thorough comprehension of total pressure drop and upstream vortex formation is paramount for optimizing the FFF process. These factors directly influence extrusion rates, backflow tendencies, and overall print quality, as previously discussed \edit{\cite{sietsepaper}}. Elevated pressure drops can give rise to complications such as filament deformation, irregular flow patterns, or even nozzle clogging. Additionally, this insight empowers us to anticipate potential challenges and develop strategies to mitigate them, thereby enhancing the efficiency and reliability of FFF-based 3D printing.

James and Roos recently demonstrated in their work \cite{james2021} that it is feasible for a viscoelastic fluid to traverse a converging channel without inducing upstream vortices, resulting in an equivalent pressure drop across the contraction as that of a Newtonian fluid with the same viscosity. Under these circumstances, elasticity does not impact pressure drop within the nozzle, but influences die swell at the exit. Achieving this involves modifying the nozzle's shape to facilitate the radial distribution of normal stress and pressure, prompting streamlines to migrate to the centerline, consequently reducing the shear rate and shear stress at the wall. Consequently, the energy expended on shear within the channel decreases, with the saved energy being redirected toward polymer-related viscous dissipation. That conclusion~\cite{james2021}, together with the result obtained in Figure~\ref{fig:vortex_vs_pressure} inspired us and allowed us to envisage the possibility of having an optimal nozzle shape for each polymer rheology so that upstream vortices would be suppressed and, consequently, the pressure drop in the nozzle would be minimum, being dependent exclusively on the viscosity of the material.


\PoF{This work showed that the tandem "nozzle shape - polymer rheology" affects the quality and performance of FFF, especially at high speeds. To prevent problems such as filament distortion, backflow and clogging, future research and development could focus on implementing Optimized Shape Design \cite{borrvall_petersson} to specific polymer rheologies. These improvements can help advance Fused Filament Fabrication technology and make 3D printing more efficient and reliable.}

\section*{Acknowledgements}
The authors thank Johan Versteegh for the fruitful discussion and selfless support. FJGR and TS acknowledge the financial support from Ultimaker B.V., and also LA/P/0045/2020 (ALiCE), UIDP/00532/2020 (CEFT) and the program Stimulus of Scientific Employment, Individual Support-2020.03203.CEECIND, funded by FEDER funds through COMPETE2020 – Programa Operacional Competitividade e Internacionalização (POCI); and by national funds (PIDDAC) through FCT/MCTES.

\FloatBarrier
\newpage
\appendix
\setcounter{table}{0}

\section[\appendixname~\thesection]{}
\label{app:1}
White~\cite{white1964} used dimensional analysis for steady second-order fluid flows, revealing three key dimensionless groups. The first, similar to classical fluid mechanics' Reynolds number ($Re = \rho U D / \mu$), represents the ratio of inertial to viscous forces, here as $Re_0 = \rho \overline{V_{ext}} D_c / \mu_0$, being $\mu_0$ the average viscosity parameter from Table~\ref{tab:prop_polymers}; the second group balances elastic and viscous forces; while the third relates to higher-order properties, particularly the ``viscoelastic ratio'' number, the ratio of $N_2$ to $N_1$, which is typically negligible in polymer melts. The group $\lambda U/L$ was termed the ``Weissenberg number,'' now known as $Wi$; the Weissenberg number used is $Wi_0 = \lambda_0  \overline{V_{ext}}  / (D_c /2)$, where $\lambda_0$ is the average relaxation time from Table~\ref{tab:prop_polymers}. Finally, the elasticity number, $El$, characterizes the elastic and inertial forces balance, defined as $El = Wi/Re$; in this document, $El_0=Wi_0/Re_0$, with $Wi_0$ and $Re_0$ as previously defined. 

Tables~\ref{tab:re0} and \ref{tab:wi0} show the values of $Re_0$ and $Wi_0$ for each material and each extrusion velocity considered in this study. It can be observed the low values of $Re_0$, below $O(10^{-5})$ and $Wi_0>1$ even for low extrusion velocities, depending on the rheological properties of the material. The small characteristic length scale and the large viscosities of every polymer melt ensure large values for $El_0$ at all the extrusion velocities.

\begin{table}[ht!]
\centering
\caption{$Re_0$ for all materials and extrusion velocities}
\resizebox{0.9\textwidth}{!}{%
\begin{tabular}{ccccccc} \toprule
 $\overline{V_{ext}}$ (mm/s)& ABS   & PLA   & PC    & PA6-66 & PET-CF & PET-G \\ \midrule
0.01 & 8.61E-11 & 1.93E-09 & 7.60E-09 & 3.11E-09   & 4.56E-09   & 6.93E-09  \\
0.1  & 8.61E-10 & 1.93E-08 & 7.60E-08 & 3.11E-08   & 4.56E-08   & 6.93E-08  \\
1    & 8.61E-09 & 1.93E-07 & 7.60E-07 & 3.11E-07   & 4.56E-07   & 6.93E-07  \\
2    & 1.72E-08 & 3.86E-07 & 1.52E-06 & 6.22E-07   & 9.12E-07   & 1.39E-06  \\
5    & 4.31E-08 & 9.64E-07 & 3.80E-06 & 1.56E-06   & 2.28E-06   & 3.47E-06  \\
10   & 8.61E-08 & 1.93E-06 & 7.60E-06 & 3.11E-06   & 4.56E-06   & 6.93E-06  \\
30   & 2.58E-07 & 5.78E-06 & 2.28E-05 & 9.33E-06   & 1.37E-05   & 2.08E-05  \\
70   & 6.03E-07 & 1.35E-05 & 5.32E-05 & 2.18E-05   & 3.19E-05   & 4.85E-05  \\
110  & 9.47E-07 & 2.12E-05 & 8.36E-05 & 3.42E-05   & 5.01E-05   & 7.63E-05  \\ \bottomrule   
\end{tabular}}
\label{tab:re0}
\end{table}

\begin{table}[ht!]
\centering
\caption{$Wi_0$ for all materials and extrusion velocities}
\resizebox{0.9\textwidth}{!}{%
\begin{tabular}{ccccccc} \toprule
 $\overline{V_{ext}}$ (mm/s)& ABS   & PLA   & PC    & PA6-66 & PET-CF & PET-G \\ \midrule
0.01 & 0.6635   & 0.01045  & 0.03565 & 0.0259      & 1.3605      & 0.01775    \\
0.1  & 6.635    & 0.1045   & 0.3565  & 0.259       & 13.605      & 0.1775     \\
1  & 66.35    & 1.045    & 3.565   & 2.59        & 136.05      & 1.775      \\
2  & 132.7    & 2.09     & 7.13    & 5.18        & 272.1       & 3.55       \\
5  & 331.75   & 5.225    & 17.825  & 12.95       & 680.25      & 8.875      \\
10   & 663.5    & 10.45    & 35.65   & 25.9        & 1360.5      & 17.75      \\
30   & 1990.5   & 31.35    & 106.95  & 77.7        & 4081.5      & 53.25      \\
70   & 4644.5   & 73.15    & 249.55  & 181.3       & 9523.5      & 124.25     \\
110  & 7298.5   & 114.95   & 392.15  & 284.9       & 14965.5     & 195.25    \\\bottomrule   
\end{tabular}}
\label{tab:wi0}
\end{table}

\FloatBarrier
\clearpage
\newpage
 \bibliographystyle{elsarticle-num} 
 \bibliography{myrefs}





\end{document}